\documentclass{ws-jbs}
\pdfoutput=1 
\usepackage{graphicx,url}
\usepackage[sort&compress]{natbib}
\begin{document}

\markboth{LW Lee, L Yin, XM Zhu, P Ao}
{Generic Enzymatic Rate Equation under Living Conditions}

\title{Generic Enzymatic Rate Equation under Living Conditions }

\author{L.~W.~Lee}
\address{Department of Mechanical Engineering,
University of Washington, Seattle, WA 98195, USA\\
\email{leelikwe@u.washington.edu}
}

\author{L.~Yin}
\address{School of Physics,
Peking University,
100871 Beijing, PR China}

\author{X.~M.~Zhu}
\address{GeneMath,
5525 27th Ave. N.E.,
Seattle, Washington 98105, USA}

\author{P.~Ao}
\address{Department of Mechanical Engineering,
University of Washington, Seattle, WA 98195, USA\\
{Department of Physics,
University of Washington, Seattle, WA 98195, USA}
\email{aoping@u.washington.edu}
}

\maketitle
\begin{history}
\received{(7 June 2007)}
\revised{(6 Aug 2007)}
\accepted{(8 Sep 2007)}
\end{history}
\begin{small}
Electronic version of an article published as {\bf Journal of Biological Systems, Vol. 15, No. 4 (2007) 495-514} [DOI:http://dx.doi.org/10.1142/S0218339007002295] © [copyright World Scientific Publishing Company] (\texttt{http://www.worldscinet.com/jbs/jbs.shtml}). Please cite the published version.
Available for download on arXiv: \texttt{http://arxiv.org/abs/0709.1696}.\\
\end{small}

\begin{abstract}
Based on our experience in kinetic modelling of coupled multiple
metabolic pathways, we propose a generic rate equation for the
dynamical modelling of metabolic kinetics. 
It is symmetric for forward and backward reactions.
It's Michaelis-Menten-King-Altman form makes the kinetic parameters (or functions) 
easy to relate to experimental values in database and to use in computation. 
In addition, such uniform form is ready to arbitrary number
of substrates and products with different stiochiometry. We explicitly show
how to obtain such rate equation rigorously for three well-known binding mechanisms. 
Hence the proposed rate equation is formally exact 
under the quasi-steady state condition.
Various features of this generic rate equation are discussed. 
In particular, for irreversible reactions, 
the product inhibition which directly arise from
enzymatic reaction is eliminated in a natural way. 
We also discuss how to include the effects of modifiers and cooperativity.
\end{abstract}


\section{Introduction}
\label{sec:introduction}

The advances in modern biology require large-scale mathematical
modelling \cite{Hood:03,Choi:07,Yang:07,Ishii:07}. One such
important task is the kinetic modelling of coupled metabolic
pathways \cite{Stephanopoulos:98,Cortassa:02,Beard:02,
Kholodenko:04,Sauro:04,Qian:05,Palsson:06,Bruggeman:06}, to understand
how an organism lives. While it is true that biological processes
are based on well-studied chemical reactions, a modeler immediately
encounters several formidable difficulties. First, kinetic equations
describing enzymatic reactions are often complicated due to numerous
parameters ~\cite{Cha:68,Segel:75}, even without considering
stochastic effects \cite{Ao:05}. Moreover, each enzymatic equation
appears different from others depending on the enzyme mechanism
involved. Therefore, modelling a simple biological process with
relatively detailed chemical and biological information can be a
daunting task \cite{Zhu:04,Kim:07}. Thus, those rigorous chemical
reactions in a coupled metabolic network do not lend themselves
easily to metabolic simulations with many reactions. Nevertheless,
many databases such as Brenda~\cite{Schomburg:02,Schomburg:04},
KEGG~\cite{Kanehisa:06}, and MetaCyc~\cite{Krieger:04,Caspi:06}
have been documenting our progress in knowledge of enzyme behavior.
If we assume that this difficulty could be overcome with great care
and effort in using the appropriate rate equations with the aid of
enzyme databases, there exists a second difficulty. Most chemical
reaction parameters are not measured under the living conditions,
how does one know they are applicable to a real organism? If not,
how could one make the appropriate and necessary adjustments in a
complicated though exact rate equation? More seriously, even if
those in vitro parameters are relevant, it is not possible to measure
all of the rate constants in the rigorous rate equation, whose number
can easily be tens of thousands for multiple coupled pathways
\cite{Storage:07,Poolman:06} in ever changing physiological conditions,
the third difficulty. The question naturally arises: Is there a way to
address those difficulties?

In this paper, based on our kinetic modelling experience we
introduce a generic rate equation capable of reproducing {\it
exactly} the full rigorous rate equations {\it irregardless} of
the enzyme mechanism for the full rate equation. The equation
takes into account thermodynamic constraints. It generalizes
easily to any reactions containing an arbitrary number of
substrates and products. The effects of modifiers (activators and
inhibitors) as well as cooperativity can be included. Thus, it
appears to be capable of addressing the above difficulties.

The outline of the rest paper is as follows. In the next section,
Sec. \ref{sec:2}, we introduce the generic enzymatic rate equation. In
Sec. \ref{sec:3}, we show how the full rigorous mechanistic rate equation
can be written {\em exactly} into the form we proposed using three
important cases as examples. In Sec. \ref{sec:4}, we discuss the case when
cooperativity is involved. In Sec. \ref{sec:5}, we discuss a useful ansatz
for the functions $f_1$ and $f_2$ and also thermodynamic
considerations in the generic equation, Eq.~(\ref{eqn:geneqn}). We
also discuss relationship to another proposed rate equation that
is different from ours and how to include the effects of
modifiers. We conclude in Sec. \ref{sec:6}.

\section{Proposed Generic Enzymatic Rate Equation }
\label{sec:2}

In this section we introduce the generic rate equation.

To further illustrate difficulties and to demonstrate usefulness
of an answer to the questions posed in the introduction, we consider the
well known Michaelis-Menten equation (see Eq.~(\ref{eqn:MM_irrev})). Once we
are dealing with bisubstrate enzymes beyond the simple Michaelis-Menten
equation~\cite{Michaelis:13} for a single substrate and product,
the rate equations becomes unwieldy due to the rapid growth in the
number of parameters in the equation. This is despite the fact
that the Michaelis-Menten equation has been shown to hold even for
a fluctuating single enzyme~\cite{English:06}. Is it possible to
fully measure the 18 terms in the steady state~\cite{Brigg:25,
Ciliberto:07} rate equation for a random Bi-Uni mechanism ({\it
c.f.} Eq.~(\ref{eqn:BiUni_full}))? The answer has been no so far
experimentally. It has been recognized that enzymes have not been
characterized to such great details in biochemical
literature~\cite{Rohwer:06}. Various rate equations in bisubstrate
enzyme reactions alone can be found in \cite{Leskovac:06}.
However, as pointed out in Ref.~\cite{Bowden:05}, the primary
concern with a rate equation used in a metabolic simulation isn't
to account for the fine difference in kinetic behavior predicted
by different mechanism. Instead, we need an equation that is
amenable to a wide variety of enzymatic reactions but is still
capable of describing the response to substrate variations and
product concentrations~\cite{Rohwer:06}. To this end, a generic
rate equation ({\it c.f.} Eq.~(\ref{eqn:geneqn})) will be very
useful since it can represent {\it exactly} the original
mechanistic rate equation. Irregardless of whether the mechanistic
details of a reaction is known, such a generic rate equation can
be adapted to all the reactions that we want to include in a
metabolic computer simulation.

Now, we come to the explicit form of the proposed rate equation.
Consider a chemical reaction with $m$ reactants, $A_1, A_2, \cdots, A_m$
and $n$ products, $P_1, P_2, \cdots, P_n$. The reaction can be
expressed as
\begin{equation}
 A_1 + A_2 + \cdots A_m \;\; ^{V_F}{\Huge\rightleftharpoons} {_{V_B}}
    \;\; P_1 + P_2 + \cdots P_n\; .
\end{equation}
Here we allow the case where some of the reactants maybe identical, i.e.
$A_i = A_j$ for some $i$ and $j$ and similarly for the products.
The reaction is assumed to be reversible, with the forward reaction
velocity $V_F$ and the backward reaction velocity $V_B$.
We propose a generic equation of the form
\begin{equation}
\displaystyle
 \nu
   = \frac{{\displaystyle V_{F} \prod_{i=1}^{m} \frac{[A_i]}{K_i}
    - V_{B}  \prod_{j=1}^{n} \frac{[P_j]}{K'_j}} }
     {{\displaystyle f_1(V_F,V_B) \prod_{i=1}^{m}
      \left(1 + \frac{[A_i]}{K_i}\right)
    + f_2(V_F,V_B) \prod_{j=1}^{n}
      \left( 1 + \frac{[P_j]}{K'_j}\right)} }  \; ,
 \label{eqn:geneqn}
\end{equation}
where $[A_i]$ with $i,\ldots,m$ and $[P_j]$ with $j,\ldots,n$ are the
concentrations of the $m$ substrates and $n$ products respectively.
Here $V_F$ and $V_B$ are the maximal forward and backward reaction velocities,
respectively. $K_i$ and $K'_j$ can be viewed as apparent Michaelis-Menten
parameter for each reactant, defining how close each reactant to saturation.
The two related functions $f_1$ and $f_2$ are introduced to take care of
production inhibition in an explicit way.
They have the following three properties:
\begin{eqnarray}
  f_1(V_F,V_B) +  f_2(V_F,V_B) & = & 1  \label{prop1}\\
  f_1(V_F=0, V_B ) & = & 0   \label{prop2}\\
  f_2(V_F, V_B=0) & = & 0  \label{prop3}\;  .
\end{eqnarray}
The first property is a normalization condition so that at very
low concentrations of $A$'s and $P$'s, the denominator is unity.
The second property express the fact that if the reaction is only
backward (i.e. $V_F$ is zero so that the first term in the
numerator vanishes), the rate should not be affected by the
concentration of $A$. Therefore, the first term in the denominator
has to be zero also. The third property has the same requirement
as the second property but applies for the case where the reaction
is irreversible in the forward direction (i.e. $V_B$ is zero so
that the second term in the numerator vanishes). In short, there
is no immediate product inhibition directly from the enzymatic reactions
for irreversible reactions where
the product does not influence $V_F$ and $K_i$. It is possible that
a reaction can be irreversible and yet also be inhibited by its
product. However, for the reactions that we discussed here, we are
concern with irreversibility in the last elementary reaction where
the enzyme product complex disassociates to give the free enzyme
and product.
We shall show clearly how irreversibility in our proposed generic rate
equation and corresponding lack of product inhibition matches exactly the
case in the full enzymatic rate equation when discussing the three cases
below. We can also model the behavior of the enzyme in situations where
product inhibition or activiation occurs in an irreversible reaction by
specifically including a product modifier term. This is discussed in
Sec.~\ref{subsec:modifiers}. 
Other properties of the proposed rate equation will be discussed
in the rest of paper.

We emphasize that $f_1, f_2 , V_F, V_B $, $\{ K_i \}$,
and $\{ K'_j \}$ should be understood as functions of
concentrations $\{ [  A_i ] \}$ and $\{ [ P_j ] \}$, and other parameters
related to their conditions in a milieu. This is important when we
show that the generic rate equation can be derived from the
mechanistic one in Sec.~\ref{sec:3}. However, since the aim is to arrive
at a generic rate equation that is convenient and suitable for modelling,
in practical situations, we can assign constant values to
$f_1, f_2 , V_F, V_B $, $\{ K_i \}$, and $\{ K'_j \}$ under certain
physiological conditions. The number of kinetic parameters needed to model
the system is greatly reduced. This is discussed further in
Sec.~\ref{subsec:5.5}.

Mathematically it may be ended here: The proposed rate equation
has already been explicitly written down. It can now be used by a
modeler in metabolic kinetic modelling. Nevertheless, the
biological implications should be further explored to give a
better understanding of the proposed generic rate equation. In the
following we will first discuss its connections to three important
enzymatic reactions \cite{Morowitz:92}:
\begin{eqnarray*}
 \mbox{(1) isomerization:}\hspace*{1.06in}
  A \mathop{\Huge\rightleftharpoons}^{V_{F\hspace*{2ex}}}_{\hspace*{2ex}V_{B}}
   P \hspace*{1.07in} \\
 \mbox{(2) binary reaction:}\hspace*{0.66in}
  A + B \mathop{\Huge\rightleftharpoons}^{V_{F\hspace*{2ex}}}_{\hspace*{2ex}V_{B}}
   P \hspace*{1.06in} \\
 \mbox{(3) condensation-splitting:}\hspace*{0.24in}
  A + B \mathop{\Huge\rightleftharpoons}^{V_{F\hspace*{2ex}}}_{\hspace*{2ex}V_{B}}
   P + Q \hspace*{0.80in}
\end{eqnarray*}
We will explicitly demonstrate that in those situations once the
rigorous kinetic equations are known our proposed equation can
exactly reproduce them. Then we discuss the connections to various
specific examples.

\section{Rigorous Derivation of Generic Rate Equation }
\label{sec:3}

In this section we show that our proposed rate equation, Eq.~(\ref{eqn:geneqn}),
is exact by explicitly discussing three important enzymatic
reactions from the classical formalisms of Michaelis-Menten or King-Altman.

\subsection{Isomerization or Uni-Uni mechanism}

We start with the best known reaction, the reversible
Michaelis-Menten mechanism (\ref{eq:uniuni}):
\begin{equation}
 E + A \mathop{\Huge\rightleftharpoons}^{k_{+1\hspace*{2ex}}}_{\hspace*{2ex}k_{-1}}
  E A  \mathop{\Huge\rightleftharpoons}^{k_{+2\hspace*{2ex}}}_{\hspace*{2ex}k_{-2}}
  E + P \, ,
 \label{eq:uniuni}
\end{equation}
where $k_{+1},k_{-1},k_{+2},k_{-2}$ are rate constants for each
elementary reaction.
It has the rate equation (see Eq.~(2.41) in~\cite{Bowden:04}, and
the explanation of each parameter there) as
\begin{equation}
 \nu = \frac{N_1 A - N_2 P}{1 + D_1 A + D_2 P}
 \label{rate:uniuni}
\end{equation}
where
\begin{eqnarray}
N_1 & = & \frac{k_1 k_2 e_0}{k_{-1} + k_2}\;;\qquad
  N_2 = \frac{k_{-1} k_{-2} e_0}{k_{-1} + k_2}\label{eqn:isoN2}\\
 D_1 & = & \frac{k_1}{k_{-1} + k_2}\;;\qquad
  D_2 = \frac{k_{-2}}{k_{-1} + k_2}
\end{eqnarray}
$e_0$ represents the total enzyme concentration.
To simplify the notation, we use $A$ instead of $[A]$ (and $P$
instead of $[P]$ ) to represent the concentration of $A$ when
writing rate equations. This simplification will be used
throughout this section (Sec. \ref{sec:3}).

Eq.~(\ref{rate:uniuni}) can be express exactly in the form of Eq.~(\ref{eqn:geneqn}), with the
constraint that $f_1 + f_2 = 1$, because Eq.~(\ref{rate:uniuni})
can be written as
\begin{equation}
 \nu = \frac{K_A N_1 \frac{A}{K_A} - K'_P N_2 \frac{P}{K'_P}}
   {f_1 \left(1 + \frac{D_1}{f_1} A \right)
  + f_2 \left(1 + \frac{D_2}{f_2} P \right) } \, ,
\end{equation}
with
\begin{eqnarray}
 K_A  & = & \frac{f_1}{D_1}\;;\qquad\quad K'_P  = \frac{1-f_1}{D_2}\\
 V_F  & = & \frac{N_1 f_1}{D_1}\;;\qquad\; V_B  =  \frac{N_2 (1-f_1)}{D_2}\;.
\end{eqnarray}

If we further choose $f_1 = V_F ^2 /( V_F ^2 + V_B ^2 )$ and $f_2
= V_B ^2 /( V_F ^2 + V_B ^2 )$ ({\it c.f.}
Eq.~(\ref{eqn:geneqn_sqdenom}) ), $K_A, K'_P, V_F$ and $V_B$ are
explicitly expressed in terms of $N_1$, $N_2$, $D_1$ and $D_2$ as
\begin{eqnarray}
 K_A & = & \frac{N_2^2 D_1}{N_1^2 D_2^2 + N_2^2 D_1^2}\;;\qquad\quad\quad
  K'_P = \frac{N_1^2 D_2}{N_1^2 D_2^2 + N_2^2 D_1^2}\\
 V_F & = & \frac{N_1 N_2^2 D_1}
           {N_1^2 D_2^2 + N_2^2 D_1^2}
           \;;\qquad\quad\quad
 V_B = \frac{N_2 N_1^2 D_2}
           {N_1^2 D_2^2 + N_2^2 D_1^2 }\;.
\end{eqnarray}
Thus, the rigorous equation of Eq.~(\ref{rate:uniuni}) is now exactly
transformed into the proposed equation, Eq.~(\ref{eqn:geneqn}), in the
isomerization case, as
\begin{equation}
 \nu = \frac{ V_{F} \frac{A}{K_A} - V_{B} \frac{P}{K'_P} }
            { f_1 \left(1 + \frac{A}{K_A}\right)
            + f_2 \left( 1 + \frac{P}{K'_P}\right)}
\label{eqn:iso}
\end{equation}

For this isomerization case, it appears there is not much to be gained in
the reduction of parameters: One starts with five parameters,
$k_{\pm 1}, k_{\pm 2}, e_0$, and ends with four, $V_{F}, V_{B},
K_A, K'_P$. We remark that the term "parameters" should be understood as 
``functions'' in the present paper, if the dependence on temperature, 
pH value, etc, are considered, which will not be explicitly spelt out here.
One would therefore wonder whether or not this would be
a pointless game. We believe it is not. In writing the rate
equation according to the proposed form, two features important in
modelling kinetics dominated by enzymatic reactions are brought
out clearly and explicitly, 
in addition to the stoichiometry encoding the molecular conservation law.

The first feature is the separation of velocities of reactions,
$V_F, V_B$, from the equilibrium or steady state concentrations
determined in most cases by Michaelis-Menton like constants, $K_A, K'_P$.
Inside a cell, though concentrations of metabolites vary according to
different physiological conditions, such variations are not very
large. On the other hand, the reaction velocities can be strongly
regulated, and can be changed over 6 orders of magnitude or
more. For example, a reversible reaction can be effectively
changed into an irreversible one. The proposed rate equation,
Eq.~(\ref{eqn:geneqn}) or Eq.~(\ref{eqn:iso}), present this feature
explicitly by separating the contributions into individual terms,
$V_F, V_B, A/K_A$ and $P/K'_P$.

The second feature is on irreversibility due to enzymatic effects.
All the concentrations are typically finite inside a cell, but the
specific product inhibition in an irreversible reaction sometimes are
not welcome.
For example, if one naively switches off the backward reaction by setting
$N_2 = 0$ in Eq.~(\ref{rate:uniuni}), there is still an apparent strong
product inhibition as the product concentration would still appear in the
denominator of right hand side of Eq.~(\ref{rate:uniuni}). The subtlety
lies in the definition Eq.~(\ref{eqn:isoN2}). If irreversibility means
the rate constant $k_{-2}$=0, then $D_2$ is zero as well so that no product
inhibition directly from this enzymatic reaction occurs. 
This is not immediately apparent to a modeller who wants
to use the rate equation Eq.~(\ref{rate:uniuni}) but vary $N_1$, $N_2$,
$D_1$ and $D_2$ independently. Nevertheless, such
apparent undesirable feature is not in the proposed rate equation. The well
known Michaelis-Menten equation for irreversible reactions without product
inhibition is readily recovered, when the backward reaction
velocity $V_B = 0$~\cite{Schnell:03,Segel:75}:
\begin{equation}
  \nu = \frac{ V_{F} A   } { K_A  + A  } \, .
\label{eqn:MM_irrev}
\end{equation}

Finally, the proposed form can be reduced back to the form for
elementary reversible reaction without enzymatic effect. This happens
when the saturation concentration is infinite, or $A/K_A \ll 1$ and
$P/K'_P \ll 1$. In this case the generic rate equation recovers the
known form at low substrate concentration:
\begin{equation}
  \nu =  \alpha_{F} A - \alpha_{B} P  \, ,
\end{equation}
with the reaction coefficients $\alpha_{F} = V_F / K_A $ and $
\alpha_{B} = V_B / K'_P $. This limiting case demonstrates that the
proposed generic rate equation is versatile.

\subsection{Binary reaction or Bi-Uni random mechanism}

\begin{figure}[ht]
 \centerline{\includegraphics[width=0.7\textwidth]{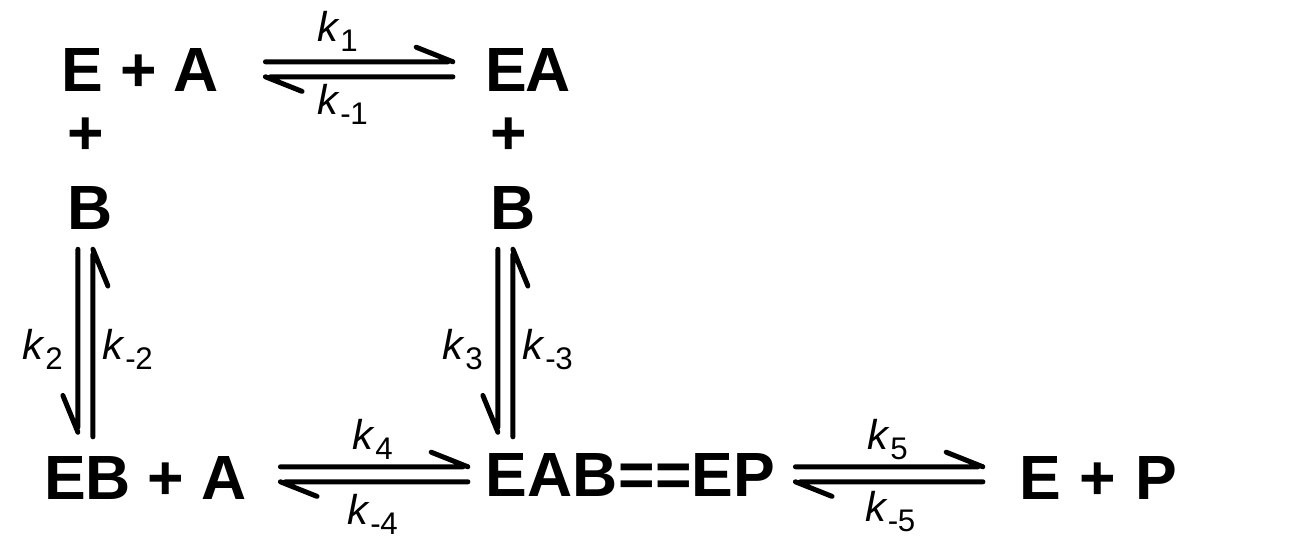}}
 \caption{Bi-Uni Mechanism\label{fig:biuni}}
\end{figure}

This is a typical reaction that occurs when two metabolites form another
one with the help of enzyme(s). A typical example of such reaction is
shown in Fig.~\ref{fig:biuni}.

The steady state rate equation for the Random Bi-Uni mechanism (see Eq.~(9.70)
in~\cite{Leskovac:03}) is
\begin{equation}
 \nu = \frac{\left(N_1 AB + N_2 A^2B + N_3 AB^2 - N_4 P - N_5 AP - N_6 BP\right)}
 {\shortstack{$\big(D_1 + D_2 A + D_3 B + D_4 AB + D_5 A^2
     + D_6 B^2 + D_7 A^2B + D_8 AB^2 $\\ \\
  $ + D_9 P + D_{10} AP + D_{11} BP + D_{12} ABP\big) $ }}
\label{eqn:BiUni_full}
\end{equation}
where the 18 terms constructed out of 11 parameters (10 rate constants:
$k_{\pm1},k_{\pm2}, k_{\pm3}, k_{\pm4}, k_{\pm5}$ and $e_0$ representing total
enzyme concentration) are
\begin{eqnarray}
 N_1 &=& (k_1 k_{-2} k_3 k_5 + k_{-1} k_2 k_4 k_5)e_0\nonumber\\
 N_2 &=& k_1 k_3 k_4 k_5 e_0\nonumber\\
 N_3 &=& k_2 k_3 k_4 k_5 e_0\nonumber\\
 N_4 &=& (k_{-1} k_{-2} k_{-3} k_{-5} + k_{-1} k_{-2} k_{-4} k_{-5})e_0\nonumber\\
 N_5 &=& k_{-1} k_{-3} k_4 k_{-5} e_0\nonumber\\
 N_6 &=& k_{-2} k_{-3} k_{-4} k_{-5} e_0\nonumber\\
 D_1 &=& k_{-1} k_{-2} k_{-3} + k_{-1} k_{-2} k_{-4} + k_{-1} k_{-2} k_{5}\nonumber\\
 D_2 &=& k_1 k_{-2} k_{-3} + k_1 k_{-2} k_{-4} + k_1 k_{-2} k_5 +
        k_{-1} k_{-3} k_4 + k_{-1} k_4 k_5\nonumber\\
 D_3 &=& k_{-1} k_2 k_{-3} + k_{-1} k_2 k_{-4} + k_{-1} k_2 k_5 +
        k_{-2} k_3 k_{-4} + k_{-2} k_3 k_5\nonumber\\
 D_4 &=& k_1 k_{-2} k_3 + k_{-1} k_2 k_4 + k_1 k_3 k_{-4} +
        k_2 k_{-3} k_4 + k_3 k_4 k_5\nonumber\\
 D_5 &=& k_1 k_{-3} k_4 + k_1 k_4 k_5\nonumber\\
 D_6 &=& k_2 k_3 k_{-4} + k_2 k_3 k_5\nonumber\\
 D_7 &=& k_1 k_3 k_4\nonumber\\
 D_8 &=& k_2 k_3 k_4\nonumber\\
 D_9 &=& k_{-1} k_{-2} k_{-5} + k_{-1} k_{-4} k_{-5} + k_{-2} k_{-3} k_{-5}\nonumber\\
 D_{10} &=& k_{-1} k_4 k_{-5} + k_{-3} k_4 k_{-5}\nonumber\\
 D_{11} &=& k_{-2} k_3 k_{-5} + k_3 k_{-4} k_{-5}\nonumber\\
 D_{12} &=& k_3 k_4 k_{-5}
\label{eqn:BiUniNandD}
\end{eqnarray}
$A,B$ and $P$ are concentrations in Eq.~(\ref{eqn:BiUni_full}), and
$E$ in Fig.~\ref{fig:biuni} represents the enzyme influence.
This is a rather complicated expression, though rigorously defined.

To make the connection to the proposed rate equation, Eq.~(\ref{eqn:geneqn}),
the numerator term can be rewritten as
\begin{equation}
  A B \underbrace{(N_1 + N_2 A + N_3 B)/D_1}_{\displaystyle V_F/(K_A K_B)}
   - P \underbrace{(N_4 + N_5 A + N_6 B)/D_1}_{\displaystyle V_B/K'_P}
\end{equation}
where the underbraces notation means we are rewriting
 ($N_1 + N_2 A + N_3 B)/D_1$ into $V_F/(K_A K_B)$ and
 $(N_4 + N_5 A + N_6 B)/D_1$ into $V_B/K'_P$.
After dividing all the terms in the denominator of
Eq.~(\ref{eqn:BiUni_full}) by $D_1$, the denominator can be written as
\begin{equation}
 \shortstack{$\big(1 + A\overbrace{(D_2 + D_5 A)}^{\displaystyle D_2}
 + B \overbrace{(D_3 + D_6 B)}^{\displaystyle D_3}
 + A B \overbrace{(D_4 + D_7 A + D_8 B)}^{\displaystyle D_4} $ \\
  $ + P D_9  + 1( D_{10} AP + D_{11} BP + D_{12} ABP)\big) $}\, .
\end{equation}
Next we introduce $f_1 + f_2 = 1$ into all ``$1$''s in the denominator
above and grouped the constants into new functions (of $A$ and
$B$) in $D_2, D_3$ and $D_4$ as shown in the overbraces:
\begin{equation}
 \shortstack{$
  f_1 + A D_2 + B D_3 + AB D_4 + f_1( D_{10} AP + D_{11} BP + D_{12} ABP)$\\
  $+ f_2 + PD_9 + f_2(D_{10} AP + D_{11} BP + D_{12} ABP)\, .$}
\end{equation}
Regrouping terms and defining $x$, $y$, $z$ and $K'_P$:
\begin{equation}
  \shortstack{$
  f_1 \Big( 1
   + A \overbrace{\left( \frac{D_2}{f_1} + D_{10} P \right) }^{\displaystyle x}
   + B \overbrace{\left( \frac{D_3}{f_1} + D_{11} P \right) }^{\displaystyle y}
   + A B \overbrace{\left( \frac{D_4}{f_1}
          + D_{12} P \right)}^{\displaystyle z} \Big) $ \\
  $ + f_2 \Big( 1 + P \underbrace{\left( \frac{D_9}{f_2} + D_{10} A
    + D_{11}B + D_{12} AB \right)}_{\displaystyle 1/K'_P} \Big) \; , $ }
\end{equation}
and introducing two functions $\alpha$ and $\beta$, we can rewrite the above as
\begin{equation}
  f_1 \Bigl(1 + A x + B y  + AB z (\alpha + \beta) + ABz(1 - \alpha -\beta)\Bigr)
  + f_2 \Bigl(1 + P/K'_P\Bigr)\;.
\end{equation}
The purpose of this is so that we can factorize into
\begin{eqnarray}
 & & f_1 \Bigl(1 + A(x + B\beta z) + B(y + A\alpha z)
  + ABz(1-\alpha-\beta) \Bigr)
  + f_2 \Bigl(1 + P/K'_P\Bigr)\nonumber\\
 & = & f_1 \Bigl(1+ A\underbrace{(x+ B\beta z)}_{1/K_A}\Bigr)
   \Bigl(1+ B\underbrace{(y + A\alpha z)}_{1/K_B}\Bigr)
  + f_2 \Bigl(1 + P/K'_P\Bigr)\;,
\end{eqnarray}
with $K_A$ and $K_B$ as defined provided
\begin{equation}
 (x + B\beta z)(y + A\alpha z) = z(1-\alpha-\beta).
\end{equation}
Note that $A, B$ and $P$ are positive being concentration and all
the $D_i$ are also positive, so $x,y$ and $z$ are positive numbers
dependent on $D_i$ and $P$. As a result, the sign of the term $B\beta z$ 
follows the sign of $\beta$. Similarly, the sign of the term $A\alpha z$
follows the sign of $\alpha$. If $x y = z$, then $\alpha=\beta=0$.
If $x y < z$, then a solution always exist with $\alpha>0$ and
$\beta>0$. If $x y > z$, then we have a solution with $\alpha<0$
and $\beta<0$. We have thus shown that
\begin{equation}
 \nu = \frac{V_F \; \frac{A}{K_A}\frac{B}{K_B} - V_B \; \frac{P}{K'_P}}
  {f_1 \; \left(1 + \frac{A}{K_A} \right)
          \left(1 + \frac{B}{K_B} \right)
   + f_2 \; \left(1 + \frac{P}{K'_P} \right)}
\label{eqn:BiUni_simple}
\end{equation}
for the binary or Bi-Uni random mechanism. This is in the form the proposed
rate equation, Eq.~(\ref{eqn:geneqn}). This form is considerably simpler in
appearance, in addition to having the explicit separation of the velocities
of reactions and steady state concentrations, which in most cases are
represented by the Michaelis-Menten like constants.
Again, for the irreversible case, if we take $k_{-5}=0$, then
$N_4$, $N_5$, $N_6$, $D_9$, $D_{10}$, $D_{11}$ and $D_{12}$ are zero in
Eq.~(\ref{eqn:BiUniNandD}). Consequently, the original rate equation
Eq.~(\ref{eqn:BiUni_full}) shows no product inhibition same as our
proposed rate equation, Eq.~(\ref{eqn:BiUni_simple}).

\subsection{Condensation-Splitting or Bi-Bi ordered mechanism}

Next we consider another well known enzymatic reaction with two substrates
and two products.
\begin{figure}[ht]
  \centerline{\includegraphics[width=0.5\textwidth]{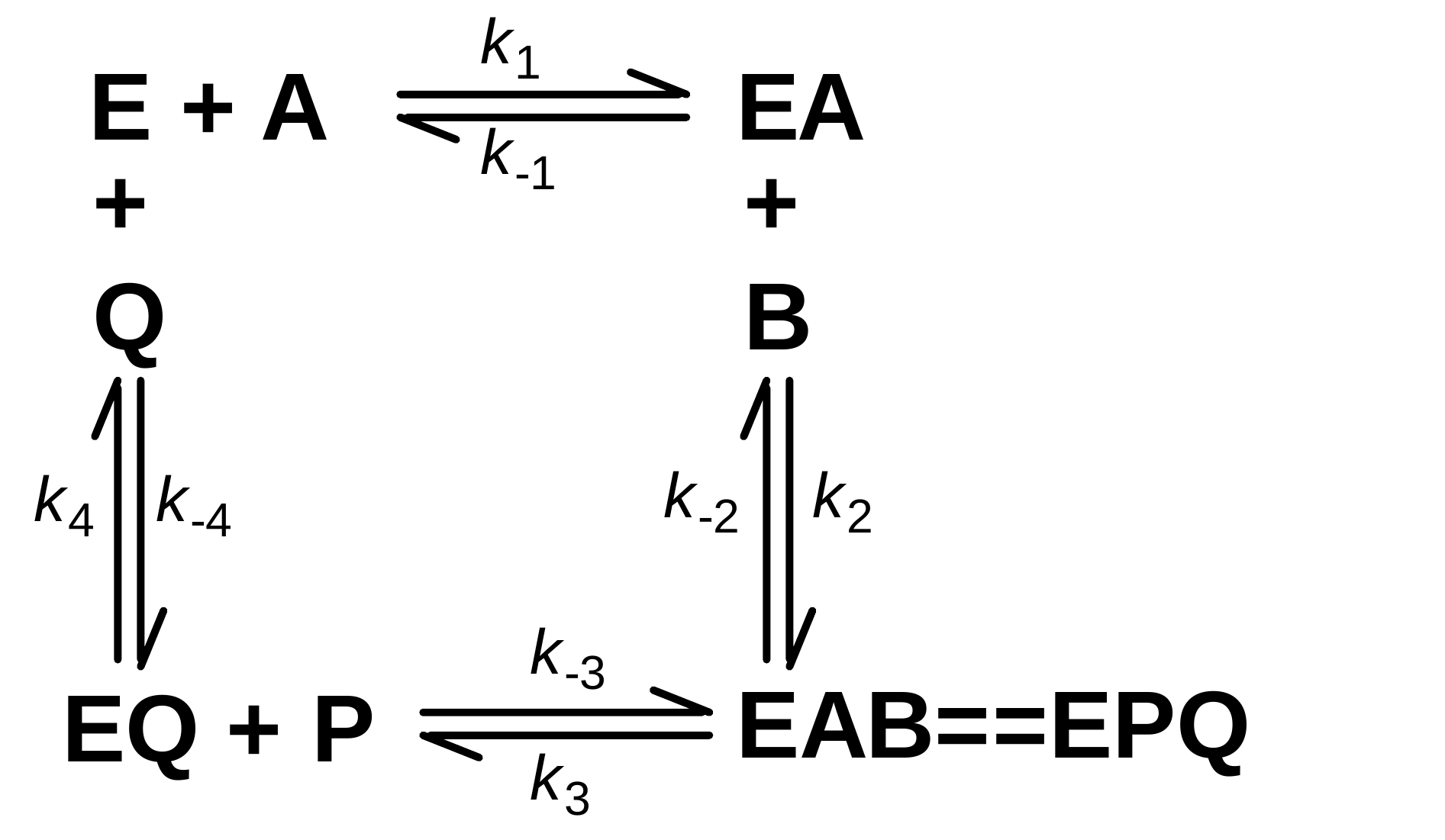}}
  \caption{Bi-Bi Ordered Mechanism\label{fig:bibi}}
\end{figure}

The steady state rate equation for the random order Bi-Bi mechanism is very
complicated with 48 terms in the denominator. The steady state rate equation
for the compulsory-order ternary-complex mechanism is simpler with 11 terms
in the denominator and is illustrated in Fig.~\ref{fig:bibi}. As shown, the
enzyme $E$ must bind first to substrate $A$ before it can react with substrate
$B$. Product $P$ is always released before product $Q$. The full rate equation
is written (see Eq.~(7.4) in~\cite{Bowden:04} or Eq.~(9.7)
in~\cite{Leskovac:03}) as
\begin{equation}
 \nu = \frac{\left(N_1 A B - N_2 P Q\right)}
  {\shortstack{$\big(D_1 + D_2 A + D_3 B + D_4 AB
  + D_5 P + D_6 Q + D_7 BQ + D_8 A P $\\ \\
  $ + D_9 PQ + D_{10} ABP + D_{11} BPQ \big).$}}
 \label{eqn:BiBi_full}
\end{equation}
The 13 terms are constructed from 8 rate constants and include $e_0$
as the total enzyme concentration:
\begin{eqnarray}
 N_1 &=& k_1 k_2 k_3 k_4 e_0\nonumber\\
 N_2 &=& k_{-1} k_{-2} k_{-3} k_{-4} e_0\nonumber\\
 D_1 &=& k_{-1} (k_{-2} + k_{3}) k_{4}\nonumber\\
 D_2 &=& k_1 (k_{-2} + k_3) k_4\nonumber\\
 D_3 &=& k_2 k_3 k_4\nonumber\\
 D_4 &=& k_1 k_2 (k_3 + k_4)\nonumber\\
 D_5 &=& k_{-1} k_{-2} k_{-3}\nonumber\\
 D_6 &=& k_{-1} (k_{-2} + k_3) k_{-4}\nonumber\\
 D_7 &=& k_2 k_3 k_{-4}\nonumber\\
 D_8 &=& k_1 k_{-2} k_{-3}\nonumber\\
 D_9 &=& (k_{-1} + k_{-2}) k_{-3} k_{-4}\nonumber\\
 D_{10} &=& k_1 k_2 k_{-3}\nonumber\\
 D_{11} &=& k_2 k_{-3} k_{-4}
\label{eqn:BiBiNandD}
\end{eqnarray}
$A,B,P$ and $Q$ are concentrations in Eq.~(\ref{eqn:BiBi_full}).

To connect such complicated equation to the proposed equation,
Eq.~(\ref{eqn:geneqn}), we first divide again the numerator and
denominator by $D_1$ and redefine the coefficients $N_1, N_2, D_2,
\ldots, D_{11}$ as divided by $D_1$.
For the numerator, we can write $N_1 = V_F/(K_A K_B)$ and
$N_2 = V_B/(K'_P K'_Q)$. For the denominator we can introduce the
functions $f_1$ and $f_2$ subject to $f_1+f_2=1$ and define
$x,y,z,x',y'$ and $z'$ as done in the Bi-Uni case to get

\begin{equation}
\shortstack{$f_1 \Big(1 + A(\overbrace{\frac{D_2}{f_1} + D_8P)}^{\displaystyle x} + B(\overbrace{\frac{D_3}{f_1}+D_7Q+D_{11}PQ)}^{\displaystyle y}
 + AB(\overbrace{\frac{D_4}{f_1} + D_{10}P)}^{\displaystyle z} \Big)$ \\ 
 + $f_2 \Big(1 + P(\underbrace{\frac{D_5}{f_2} + D_8A + D_{10}AB)}_{\displaystyle x'} + Q(\underbrace{\frac{D_6}{f_2} + D_7B)}_{\displaystyle y'}
 + PQ(\underbrace{\frac{D_9}{f_2} + D_{11}B)}_{\displaystyle z'} \Big)\, .$}
\end{equation}

By introducing two functions $\alpha$ and $\beta$ for the $f_1$
bracket part and two functions $\alpha'$ and $\beta'$ for the
$f_2$ bracket part, we can use the same method as in the Bi-Uni
case to rewrite the denominator into
\begin{equation}
   f_1 \left(1 + \frac{A}{K_A} \right)\left(1 + \frac{B}{K_B} \right)
 + f_2 \left(1 + \frac{P}{K'_P}\right)\left(1 + \frac{Q}{K'_Q}\right)\, .
\end{equation}
The rate equation is then
\begin{equation}
 \nu = \frac{V_F \; \frac{A}{K_A}\frac{B}{K_B} - V_B \; \frac{P}{K'_P}\frac{Q}{K'_Q}}
   {f_1 \; \left(1 + \frac{A}{K_A} \right)\left(1 + \frac{B}{K_B} \right)
  + f_2 \; \left(1 + \frac{P}{K'_P}\right)\left(1 + \frac{Q}{K'_Q}\right)}
  \label{eqn:BiBi} \; ,
\end{equation}
precisely in the form of the proposed equation, Eq.~(\ref{eqn:geneqn}).
Furthermore, if the product releasing steps are irreversible, i.e. $k_{-3}=0$
and $k_{-4}=0$, then $N_2$, $D_5$, $D_6$, $D_7$, $D_8$, $D_9$, $D_{10}$ and
$D_{11}$ are zero in Eq.~(\ref{eqn:BiBiNandD}). Hence there is no product
inhibition in original rate equation, Eq.~(\ref{eqn:BiBi_full}) same as our
proposed form, Eq.~(\ref{eqn:BiBi}) when irreversible.
Now, there should be no question that the proposed equation is simpler,
though formally it is equivalent to the rigorous equation as shown.

\section{Cooperativity for the case ${\ } n \; A \mathop{\rightleftharpoons} P$ }
\label{sec:4}

In this section, we discuss how to include cooperative
effects~\cite{Sorribas:07} in enzyme.
Hill's equation is often used in enzymatic kinetics to account for
cooperativity. It came from early attempt by Archibald Hill to
describe the sigmoidal $O_2$ binding curve of haemoglobin in
1910~\cite{Hill:10}.
If the binding of ligand by a protein with several sites can be
adequately described by the following equation
\begin{equation}
  E + nA \mathop{\Huge\rightleftharpoons} E(A)_n
\end{equation}
then the equilibrium expression is given by
\begin{equation}
  K_d = \frac{[E][A]^n}{[E(A)_n]}
  \quad \mbox{or}\quad [E(A)_n] = \frac{[E][A]^n}{K_d}
  \label{eqn:hillequi}
\end{equation}
where $K_d$ is the disassociation constant. The
fraction of ligand binding sites filled is
\begin{equation}
 Y = \frac{[E(A)_n]}{[E] + [E(A)_n]}\, .
\end{equation}
Substituting in Eq.~(\ref{eqn:hillequi}), we get
\begin{equation}
 Y = \frac{[A]^n}{K_d + [A]^n}\, .
\end{equation}
We can define $K_d = K_{0.5}^n$ so that $K_{0.5}$ has units of concentration
and when $K_{0.5} = [A]$, half of the binding sites are filled.
The reaction rate for the Hill equation~\cite{Hill:10} in the irreversible
case then follows
\begin{equation}
\nu = \frac{V_F [A]^n}{K_{0.5}^n + [A]^n}\, .
\end{equation}

This can be generalized to the reversible case~\cite{Bowden:97}
\begin{equation}
  \nu = \frac{\left(V_F \frac{[A]}{A_{0.5}} - V_B \frac{[P]}{P_{0.5}}\right)
 \left(\frac{[A]}{A_{0.5}} +  \frac{[P]}{P_{0.5}}\right)^{h-1}}
 {1 + \left(\frac{[A]}{A_{0.5}} +  \frac{[P]}{P_{0.5}}\right)^h}\;,
\end{equation}
where $h$ is the Hill's coefficient.
In our case, the proposed generic rate equation would be written as
\begin{equation}
 \nu = \frac{V_F (\frac{[A]}{K_{A}})^h - V_B \left(\frac{[P]}{K'_{P}}\right)}
   {f_1(V_F, V_B)\left( 1 +  \left(\frac{[A]}{K_{A}} \right)^h \right)
  + f_2(V_F, V_B) \left(1 + \frac{[P]}{K'_{P}}\right)}\;.
\end{equation}
Cooperativity is included by raising $A/K_A$ to a higher power. If desired,
we can also raise $P/K'_P$ to a higher power if cooperativity is present in
the reverse direction. Using the same method of grouping terms and
introducing additional functions in the previous section (Sec. \ref{sec:3}),
it should be possible to rewrite our proposed equation to the desired
cooperative form.

\section{Discussions}
\label{sec:5}

\subsection{Ansatz for $f_1$ and $f_2$}

Under the conditions of Eqs.~(\ref{prop1}--\ref{prop3}),
a useful ansatz for the functions $f_1$ and $f_2$ would be
\begin{eqnarray}
  f_1(V_F,V_B) & = & \frac{\delta V_F ^{\gamma_1} }
   { \delta V_F^{\gamma_1} + \epsilon V_B ^{\gamma_2} }  \label{eqn:ansa1}\; ,\\
  f_2(V_F,V_B) & = & \frac{\epsilon  V_B ^{\gamma_2} }
   { \delta V_F ^{\gamma_1} + \epsilon V_B ^{\gamma_2} } \label{eqn:ansa2}\; ,
\end{eqnarray}
with positive numerical constants $\delta, \epsilon, \gamma_1$ and
$\gamma_2$. The relative weight of forward and backward reactions
are determined by $\delta$ and $\epsilon$, while $\gamma_1$ and
$\gamma_2$ determine the relative influence of $V_F$ and $V_B$ to
the generic rate equation, Eq.~(\ref{eqn:geneqn}). The larger the
$\gamma$'s are, the smaller their contribution to the denominator
when the reaction velocity is small. When $\gamma$'s are infinite,
there is no contribution from the smaller reaction. For example,
if $\gamma_{1}=\gamma_{2} = \infty$ and $V_F > V_B$, then $f_1 = 1$
and $f_2 = 0$. On
the other hand, the smaller the $\gamma$'s are, the larger
relatively the influence from the small reaction velocity. When
$\gamma$'s are zero, the denominator is independent of the
reaction velocities: $f_1(V_F,V_B)  =  \delta /( \delta + \epsilon
),  f_2(V_F,V_B) =  \epsilon / ( \delta + \epsilon )$.

If we wish the relative effect of smaller reaction velocity
to be on the smaller side, the numerical parameters in
Eqs.~(\ref{eqn:ansa1}--\ref{eqn:ansa2}) may be chosen as
$\gamma_1 = \gamma_2 = 2$.
We further take $\delta =\epsilon = 1$ so that forward and backward
reactions are equally important. The reaction rate would then be
generally written as
\begin{equation}
 \nu
  = \frac{{\displaystyle
           V_{F} \prod_{i=1}^{m} \left(\frac{[A_i]}{K_i}\right)
         - V_{B} \prod_{j=1}^{n} \left(\frac{[P_j]}{K'_j}\right)}}
    {{\displaystyle \frac{ V_F ^2 }{ V_F ^2 + V_B ^2 }
     \prod_{i=1}^{m} \left(1 + \frac{[A_i]}{K_i}\right)
   + \frac{ V_B ^2 }{V_F ^2 + V_B ^2 }
     \prod_{j=1}^{n} \left(1 + \frac{[P_j]}{K'_j}\right)}}
\label{eqn:geneqn_sqdenom}
\end{equation}
For irreversible reactions, say $V_B=0$, the effect of the products
concentrations drops out naturally from the denominator.
Eq.~(\ref{eqn:geneqn_sqdenom}) can be generalized to arbitrary reactions.

\subsection{Thermodynamics }

We next discuss how the generic rate equation,
Eq.~(\ref{eqn:geneqn}) also satisfy thermodynamic constraints. The
equilibrium constant is
\begin{equation}
 K_{\rm eq} = \frac{\displaystyle \prod_{j=1}^{n} [P_j]}
                   {\displaystyle \prod_{i=1}^{m} [A_i]}
 \label{eqn:keq}
\end{equation}
At equilibrium, the reaction rate is zero. Hence $K_{\rm eq}$ can
also be expressed as
\begin{equation}
 K_{\rm eq} = \frac{{\displaystyle V_F \prod_{j=1}^{n} K'_j}}
                   {{\displaystyle V_B \prod_{i=1}^{m} K_i}}
\end{equation}
(also called Haldane equation). Consequently, we can write the
generic rate equation, Eq.~(\ref{eqn:geneqn}) as
\begin{equation}
\nu
  = \frac{\displaystyle \left(V_{F} \prod_{i=1}^{m}\frac{[A_i]}{K_i} \right)
     \left(1 - \frac{\Gamma}{K_{\rm eq}}\right)}
{{\displaystyle f_1(V_F,V_B) \prod_{i=1}^{m}\left(1 + \frac{[A_i]}{K_i}\right)
              + f_2(V_F,V_B) \prod_{j=1}^{n}\left(1 + \frac{[P_j]}{K'_j}\right)} }
\end{equation}
where
\begin{equation}
\Gamma = \frac{\displaystyle\prod_{j=1}^{n} [P]}
              {\displaystyle\prod_{i=1}^{m} [A]}
\end{equation}
is often called the mass action ratio. It can be seen that the
equation obeys thermodynamics, since when $\Gamma = K_{\rm eq}$,
the reaction rate is zero. When $\Gamma > K_{\rm eq}$, the
reaction is in the reverse direction, which restore the
equilibrium. For $\Gamma < K_{\rm eq}$, the rate is in the forward
direction.

It should be pointed out that under living conditions, the
medium in which reactions take place is active, that is, it is an
open system. Since the reactions are usually very complicated,
often not all reactants are written out explicitly. For example,
some abundant reactants such as water ($H_2 O$) and carbon
dioxides ($C O_2$) may be omitted from a reaction. In addition,
some tightly regulated metabolites, such as adenosine
triphosphate ($ATP$) may be regarded as constant through out the
processes under consideration, and hence may be omitted from the
reaction. Because of such active and open nature under living
conditions, simple thermodynamical consideration may be
misleading simply because the system is not in an equilibrium typically 
defined for a closed system.

\subsection{Comparison with other proposals}

The need to use simple (and even approximated~\cite{Heijnen:05,Wang:07})
rate equations have
been noted for some time~\cite{Bowden:05}. An equation different
from ours has been proposed for the bi-substrate
case~\cite{Bowden:05,Rohwer:06}:
\begin{equation}
 \nu
  = \frac{ \left( 1 - \frac{ [P][Q]}{K_{\rm eq}[A][B]}\right) \frac{V[A][B]}{K_A K_B} }
   { \left(1 + \frac{[A]}{K_A} + \frac{[Q]}{K'_P}\right)
     \left(1 + \frac{[B]}{K_B} + \frac{[P]}{K'_Q}\right) }  \; .
 \label{eq:bowdeneq}
\end{equation}
$K$ may be viewed as apparent Michaelis constant with one
product present. For example, $K_A$ could be estimated as the
apparent Michaelis constant for $A$ with $B$ and $P$ at
appropriate concentrations and $Q$ absent~\cite{Bowden:05}.

Comparing Eq.~(\ref{eq:bowdeneq}) with our propose
Eq.~(\ref{eqn:geneqn}), the first apparent difference is in the
numerator.  However, a close inspection of the rate equation,
Eq.~(\ref{eq:bowdeneq}) indicates that the denominator may
overestimates the contribution from the backward reaction. For
example, when the backward reaction velocty
 is zero, we should
not expect that the denominator will be directly affected by the
product: it is irreversible. The equation we proposed,
Eq.~(\ref{eqn:geneqn}) does not contain this problem since $f_2$ is
zero whenever $V_B$ is zero.

Another point to note is that the derivation of Eq.~(\ref{eq:bowdeneq})
depends on the binding scheme used.
Two different formulations can be obtained depending on whether
substrate $A$ and product $P$ binds to one site and $B$ and $Q$ binding
to the second site, or whether $A$ pairs with $Q$ and $B$ with $P$
\cite{Rohwer:06}. This requires some knowledge of the enzyme binding
mechanism which diminish it's usefulness as a generic equation.

\begin{figure}[ht]
 \centerline{\includegraphics[width=1.0\textwidth]{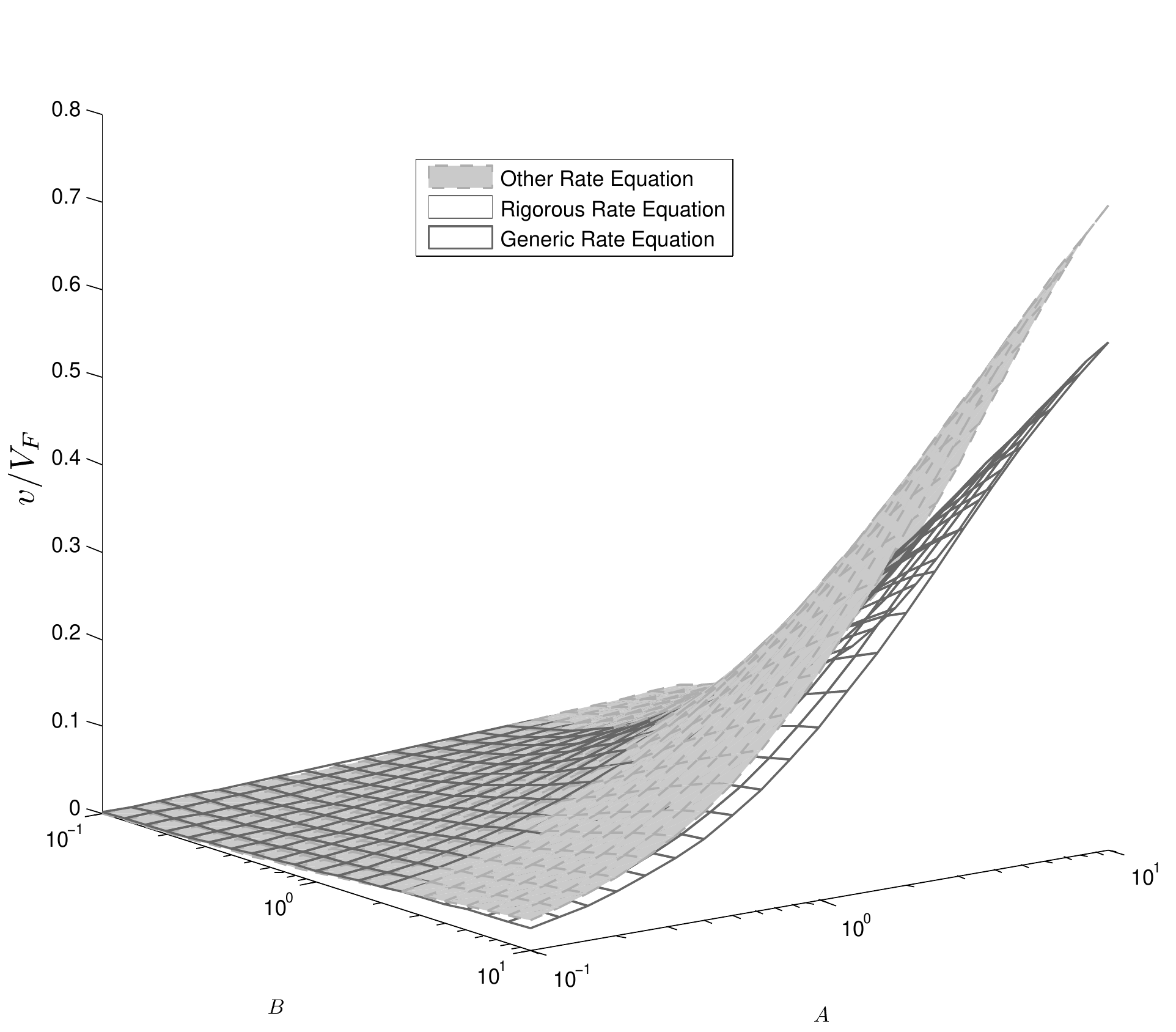}}
 \caption{Comparison of different rate equations. Here we take the Bi-Bi
 ordered mechanism with 8 rate constants as an example. In
 Ref.~\cite{Rohwer:06}, a generic bi-substrate rate equation was fitted to
 the mechanistic rate equation according to a set of data points for the
 substrates and products concentrations~(Conditions CIII in \cite{Rohwer:06},
 see their Fig. 2).
 The fit at product concentration $P=Q=0.1$ is considered good. Nevertheless,
 our proposed generic rate equation, Eq.~(\ref{eqn:geneqn}), is capable of
 reproducing {\it exactly} the original rigorous mechanistic rate equation.
 We note that it is likely the other equation, Eq.~(\ref{eq:bowdeneq}) may
 also be cast into the rigorous rate equation using the method shown in
 Sec. \ref{sec:3}.
 \label{fig:trymesh} }
\end{figure}

The proposed generic rate equation in Ref.~\cite{Rohwer:06} was compared
to the original mechanistic rate equation in the case of Bi-Bi ordered
mechanism. The fit was found to be good under some conditions
(see Fig.~\ref{fig:trymesh}).
In contrast, our generic rate equation can represent {\it exactly} the
original mechanistic rate equation as illustrated in Fig.~\ref{fig:trymesh}.

\subsection{Adding modifiers}
\label{subsec:modifiers}
Modifiers are substances that interact with enzymes to either
increase or decrease their catalytic activity. Inhibitors diminish
the rate of reaction while activators increase the rate. These are
very important in the regulation of metabolic pathways in
organisms. For reversible inhibitors, they can be classified into
competitive inhibitors and noncompetitive inhibitors. Competitive
inhibitors compete with the substrate for binding to the enzyme
forming an enzyme-inhibitor complex. The inhibitor does not act on
the enzyme-substrate complex once it is formed. For the simple
Michaelis-Menten kinetics, the resulting form is
\begin{equation}
 \nu = \frac{V_F [A]}{K_A \left(1 + \frac{[I]}{K_I} \right) + [A]}\, .
 \label{eq:competitive}
\end{equation}
$V_F$ is not affected but $K_A$ is increased.
Non-competitive inhibitors bind with the same affinity to the enzyme
and enzyme-substrate complex. It does not change $K_A$ but decrease
$V_F$:
\begin{equation}
 \nu = \frac{ \left(\frac{1}{1 + {[I]}/{K_I}} \right) V_F [A]}{K_A + [A] }\ .
 \label{eq:non-comp}
\end{equation}
When the affinities are different, the inhibition is called mixed.
The net result is to increase in $K_A$ and decrease in $V_F$.
A special case occurs if inhibitor only binds to the
enzyme-substrate complex but not to the enzyme, this is called
uncompetitive:
\begin{equation}
 \nu =
   \frac{ \left(\frac{1}{1 + [I]/K_I} \right) V_F [A]}
        { \left(\frac{1}{1 + [I]/K_I} \right) K_A + [A] }\, .
\end{equation}

To include the effects of inhibitors and activators, we can multiply
the unmodified rate equation with a function of the sigmoidal form
\begin{equation}
 \frac{1}{1 + ([I]/a)^2}\,,
\label{eq:inhib}
\end{equation}
for an inhibitor, $I$ and
\begin{equation}
 \frac{([T]/b)^2}{1 + ([T]/b)^2}\,,
\end{equation}
for an activator, $T$. The parameter $a$ can be adjusted to
match the concentration of the inhibitor at which inhibition
becomes significant. The same applies for $b$ in the case of
an activator. Other power in the denominator besides 2 can also be
used. It is clear that when $V_F$ is modified by multiplying the
modifier function, this is similar to the case for non-competitive
inhibitor in Eq.~(\ref{eq:non-comp}).
For the competitive inhibitor case, a suitable choice of
$1/K_{I} = (1+\frac{[A]}{K_A})/a$ substituted into Eq.~(\ref{eq:competitive})
gives
\begin{eqnarray}
\nu & = & \frac{{V_F}[A]}{K_A \left(1 + [I]/K_I\right) + [A]}\nonumber\\
	& = & \frac{V_F [A]}{K_A \left(1 + ([I]/a) (1 + [A]/K_A)\right) + [A]}\nonumber\\
	& = & \frac{V_F [A]}{(1 + [I]/a)(K_A + A)}.
\end{eqnarray}
This is the same form of the inhibitor that we used in Eq.~(\ref{eq:inhib}).
Hence we have shown that both competitive and non-competitive modifiers
can generally be accounted for by the modifier form that we used in
Eq.~(\ref{eq:inhib}).

\subsection{Uses of proposed rate equation}
\label{subsec:5.5}

We have demonstrated above that in situations where exact enzymatic
reactions are known, the proposed rate equations are the same as them. 
In realistic modelling situations, the enzymatic information is typically
incomplete. The proposed equations are easy to use because of its simplicity
and its parameters or functions are easy to interpret. 
We elaborate on this observation further below.

Earlier in deriving the generic rate equation (in Sec.~\ref{sec:3}),
we take $V_F$, $V_B$ and, $K_i$'s where $i$ are metabolites in the
enzymatic reaction, to be
functions of concentrations of metabolities and rate constants in
the elementary reactions. We already alluded to the situation of incomplete
information, and to the nearly impossibility of obtaining all the relevant
{\it in vivo} kinetic parameters experimentally.
Fortunately, it is well known that the biological networks, either genetic or
metabolic, or others, are typically robustly built
\cite{Choi:07, Stephanopoulos:98, Cortassa:02, Kholodenko:04, Palsson:06,
Zhu:04, Kim:07} at functional level.
Therefore, very detailed kinetic information may not be necessary in practical
situations where we want to simulate a metabolic model under some
physiological or functional conditions. We may assign typical values, 
which may be approximated by constants, 
to the parameters $V_F$, $V_B$ and $K_i$ in the simulation. 
Since the operational meaning of these parameters are clear
(see below Eq.~(\ref{eqn:geneqn})), we can adjust
these parameters in a meaningful way to match physiological conditions.

There is another way to use the proposed enzymatic equation:
The compact form of Eq.~(\ref{eqn:geneqn}) with approximated values in 
the kinetic parameters may serve as a starting point to introduce empirical
description. Once the values of parameters are fixed, 
specific experiments to measure them can be performed, 
and the comparison between theoretical and experimental
efforts can lead to better understanding and further study.
Clearly, this appears a more straightforward procedure than 
tuning the rate constants of the elementary reactions 
(such as the $k_i$'s in Eq.~(\ref{eqn:BiUniNandD})) in
the original rate equation (such as Eq.~(\ref{eqn:BiUni_full})).
Effectively, the number of parameters in the model is reduced. 
Can such a strategy be useful? Our experience so far suggests yes.

From a different angle, a third possible use of the proposed enzymatic
equation is to regard
it as a format to document kinetic parameters from experiments: The form
is formally exact, with easy to interpret parameters, and, is ready
for computational modeling. In particular, it would be of great value
if experimental results on reversible reactions, 
particularly on the backward velocities, are documented.

We have been using the generic rate equation,
Eq.~(\ref{eqn:geneqn_sqdenom}) to construct a kinetic model of the
metabolism of \textit{Methylobacterium
extorquens AM1}~\cite{Christoserdova:03,Dien:02,Dien:03} consisting of
about 80 reactions and 80 metabolites. The number of metabolites are
the minimum in order to understand the metabolic kinetics at a global
level~\cite{Christoserdova:03,Downs:06}. We are able to find a set
of parameters ($V_F$, $V_B$ and $K_i$'s) that give a steady state
solution \cite{Yin:07}. Our results indicate that the model is robust with
respect to variations in kinetic parameters. This is in agreement with
observations that quite large errors in kinetic parameters for
most of the enzymes have very little effect on the calculated
flux~\cite{Bowden:05}. This result is very encouraging as it
suggest even larger metabolic network can be built on the same
structure for kinetic equations.

Reducing the number of kinetic parameters are also useful when
metabolomic data are available for analysis. Under such situation,
one could attempt to fit the kinetic parameters to the metabolomic
data. One way may be to use Genetic Algorithms to optimize the
parameters~\cite{Nobuyoshi:05}. Simplifying the number of kinetic
parameters cuts the computational effort and makes the problem
much more tractable.

\section{Conclusions}
\label{sec:6}

We have proposed a generic kinetic rate equation. It is shown to
be formally identical to rigorous equations and has a
considerably simple and generic form. It is already in a form
workable for arbitrary number of substrates and products with
different stiochiometry. Modifier functions can be added to
reflect inhibition and activation. Cooperativity can also be
included.

In order to do kinetic modelling according to rigorous equations,
numerous parameters need to be measured. Instead, the proposed
generic rate equation reduces the required number to the most
essential ones, both forward and backward reaction velocities,
$V_F$, $V_B$, and the Michaelis-Menten like constants, $K, K'$,
for both substrates and products, which have clear chemical and
biological meanings. Even if those essential parameters are not
always available under appropriate physiological conditions
related to living organisms in databases, their adjustments
during modelling are much more managable.

The kinetic parameters from enzyme characterization is also
dependent on the rate equation being tested. There is an effort
underway to ensure uniformity in reporting functional
data~\cite{Apweiler:05}. The proposed rate equation with only essential
parameters may aid such effort to have enzyme characterized. With a
database of enzyme parameters for a generic rate equation of the
form Eq.~(\ref{eqn:geneqn}), the resulting uniformity will make
reconstruction of genome scale metabolic networks using kinetic
models will become much more feasible.

\section*{Acknowledgments}
 We acknowledge theoretical and mathematical discussions with
 James B. Bassingthwaighte,
 George Kos{\'a}ly, Hong Qian, and Steve van Dien, and inputs which have greatly
 enhanced our biological understanding from
 Greg Crowther, Sergei Stolyar, Mary Lidstrom and the entire members of her lab.
 This research was supported in part by NIH under grant \# GM36296 (LWL) and
 under grant \# HG002894 (LY and PA).

\bibliography{refs}

\end{document}